\documentclass[12pt]{article}
\usepackage{epsf}
\usepackage{latexsym}
\newcommand{\be}{\begin{equation}}
\newcommand{\ee}{\end{equation}}
\newcommand{\ba}{\begin{array}}
\newcommand{\ea}{\end{array}}
\newcommand{\bqa}{\begin{eqnarray}}
\newcommand{\eqa}{\end{eqnarray}}
\begin{document}
%%%%%%%%%%%%%%%%%%%%%%%%%%%%%%%%%%%%%%%%%%%%%%%%%%%%%%%%%%%%%%%%%%
%%%%%%%%%%%%%%%%%%%%%%%%%%%%%%%%%%%%%%%%%%%%%%%%%%%%%%%%%%%%%%%%%%%
\begin{center}
{\Large\bf A Critical Examination to the Unitarized   $\pi\pi$
Scattering Chiral Amplitudes}
\\[10mm]
{\sc  Qin Ang, Zhiguang Xiao, H.~Zheng and X.~C.~Song}
\\[5mm]
{\it Department of Physics, Peking University, Beijing 100871,
P.~R.~China }
\\[5mm]
\today
\begin{abstract}
We discuss the Pad\'e approximation to the $\pi\pi$ scattering
amplitudes in 1--loop chiral perturbation theory. The
approximation restores unitarity and can reproduce the correct
resonance poles, but the approximation violates crossing symmetry
and produce spurious  poles on the complex $s$ plane and therefore
plagues its predictions on physical quantities at quantitative
level. However we find that one virtual state in the IJ=20 channel
may have physical relevance.
\end{abstract}
\end{center}
\vspace{1cm}

The chiral perturbation theory~\cite{GL}  is a powerful tool in
studying strong interaction physics at low energies. The principle
of chiral perturbation theory is that it incorporates the global
symmetry of the QCD Lagrangian -- the spontaneously broken chiral
symmetry -- into the effective Lagrangian, and directly deals with
the physical states -- the pseudo-Goldstone bosons -- as the
physical degrees of freedom in the effective Lagrangian. The
physical amplitude is calculated by a perturbative expansion in
powers of the light quark masses and the external momentum, and
unitarity is respected perturbatively. As the external momentum
increases, however, the chiral expansion diverges  rapidly and the
unitarity condition  is badly violated. To remedy such a
situation, the Pad\'e approximation has been used to improve the
behavior of chiral amplitudes at higher energies, which has
stimulated revived interests in the recent literature (see for
example \cite{truong} --\cite{OOR00}). One of the cost in using
Pad\'e approximation
 is the violation of crossing
symmetry which has also been discussed in the recent literature
(see for example Refs.~\cite{BP97,CB01} and \cite{DP97}).

For the $\pi\pi\rightarrow \pi\pi$ scattering process, the isospin
amplitudes in the s channel can be decomposed as~\cite{MMS},
\bqa
T^{I=0}(s,t,u) & = & 3A(s,t,u)+A(t,u,s)+A(u,s,t)\ ,\nonumber \\
T^{I=1}(s,t,u) & = & A(t,u,s)-A(u,s,t)\ , \\
T^{I=2}(s,t,u) & = & A(t,u,s)+A(u,s,t)\ , \nonumber
\eqa
as a  result of the generalized Bose statistics and crossing symmetry.
In $SU(2)\times SU(2)$ chiral
perturbation theory (ChPT) to one loop~\cite{GM} we have,
\bqa A(s,t,u)&=&{s-m_\pi^2\over
F_\pi^2}+A_1(s,t,u)+A_2(s,t,u)+O(E^6), \nonumber \\
A_1(s,t,u)&=&{1\over 6F_\pi^4}\{3(s^2-m_\pi^4){\bar J}(s)
+\left[t(t-u)-2m_\pi^2t+4m_\pi^2u-2m_\pi^4\right]{\bar
J}(t)\nonumber \\ &&
+\left[u(u-t)-2m_\pi^2u+4m_\pi^2t-2m_\pi^4\right]{\bar
J}(u)\},\nonumber\\ A_2(s,t,u)&=&{1\over 96\pi^2F_\pi^4}\{2(\bar
l_1-{4\over3})(s-2m_\pi^2)^2 +(\bar
l_2-{5\over6})\left[s^2+(t-u)^2\right]-12m_\pi^2s\nonumber \\ &&
+15m_\pi^4+12m_\pi^2(s-m_\pi^2) \bar l_4-3m_\pi^4\bar l_3\}
 \eqa
where $F_\pi=93.3$MeV is the pion decay constant and the function
${\bar J}(s)$ is defined as \bqa {\bar J}(s) &=& {1\over
16\pi^2}\left [\rho{\rm ln}\left({\rho-1\over
\rho+1}\right)+2\right ] , \nonumber \\ \rho(s) & =
&\sqrt{1-{4m_\pi^2\over s}}\ . \eqa In Ref.~\cite{GM} the $\bar
l_i$ parameters are taken to be \be\label{sd} \bar
l_1=-0.62\pm0.94 \ , \,\,\, \bar l_2=6.28\pm0.48\ ,\,\,\, \bar
l_3=2.9\pm2.4\ ,\,\,\, \bar l_4=4.3\pm0.9 \ee determined from low
energy experiments.

The partial wave expansion of the isospin amplitudes is written as
\bqa T^I(s,t,u)&=& 32\pi \sum_J(2J+1)P_J(\cos\theta){T^I_J}(s)\ ,
\eqa where, for $\pi\pi$ scatterings, the sum is over even(odd)
values of $J$ for even(odd) values of $I$ because of the
restriction of Bose statistics. The inverse expression is \bqa
 T_J^I(s) &=& {1\over 64\pi} \int^{1}_{-1}
d{\cos\theta}\, P_J(cos\theta) T^{I}(s,t,u)\ ,\nonumber \\
 \cos\theta &=& 1+{{2t}\over{s-4m_\pi^2}} \ , \\
u &=& 4m_\pi^2-s-t\ . \nonumber
\eqa
The partial wave amplitudes in ChPT  expanded to
$O(p^4)$ are,
\be T^I_J(s) = T^I_{J, 2}(s)+T^I_{J,4}(s)\ ,
\ee
where $T^I_{J,2}(s)$ and $T^I_{J,4}(s)$ represent
terms of order $O(p^2)$ and $O(p^4)$, respectively. The $T^I_{J,2}(s)$
amplitudes can be rigorously derived from current algebra and are model independent.
The  [1,1] Pad\'e approximation to the partial
wave amplitudes is,
\be\label{pade}
T^{I[1,1]}_J(s) = {{T^I_{J,
2}(s)}\over{1-{T^I_{J, 4}(s)/T^I_{J, 2}(s)}}}\quad .
\ee
From perturbative
unitarity in ChPT we have
\bqa\label{PU}
{\rm Im}T^I_{J,
4}(s)&=&\rho|T^I_{J, 2}(s)|^2\ ,\,\, (4m_\pi^2<s<16m_\pi^2)\ .
\eqa
With this relation, it is easy to prove that the [1,1]
Pad\'e approximants given in Eq.~(\ref{pade}) satisfy elastic unitarity:
\be
{\rm Im}T^{I[1,1]}_J(s) = \rho|T^{I[1,1]}_J(s)|^2 \quad ,
\ee
and it is well known that
the phase shifts they predict are considerably improved comparing with
the perturbation results. However,
the total amplitude
 \be T(s,t,u)= 32\pi
\sum_J(2J+1)P_J(cos\theta){T_J}(s)\ ,
\ee
has  certain crossing
properties which are lost in the amplitude defined by
 \be
T^P(s,t,u)\equiv 32\pi
\sum_J(2J+1)P_J(cos\theta){T^{[1,1]}_J}(s)\ ,
\ee
by violating the so called  Balachandran-Nuyts-Roskies  relations,
as discussed recently in Refs.~\cite{BP97,CB01}.

\begin{figure}[hbtp]
\begin{center}
\vspace*{0mm}
\epsfysize=70mm
\epsfbox{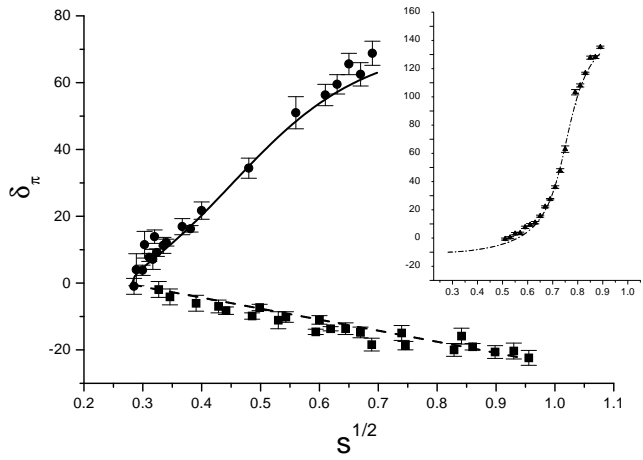}
\vspace*{5mm} \caption{ \label{phase} The global fit
in  IJ=00, 11 and 20 channels,
using the [1,1] Pad\'e approximants and the procedure described in the text.
}
\end{center}
\end{figure}

The Pad\'e approximation not only gives an improved prediction to
the $\pi\pi$ scattering phases comparing to the perturbative
results, but also it predicts the correct physical resonances on a
qualitative level. For example in the amplitude $T^{0[1,1]}_0(s)$,
on the second sheet of the complex $s$ plane one finds the
$\sigma$ resonance with $M_\sigma=430$MeV and
$\Gamma_\sigma=456$MeV,~\footnote{Using the same set of $\bar l_i$
parameters as in Eq.~(\ref{sd}) and the  method of
Ref.~\cite{XZ00}, a  fit to the experimental phase shift in the
IJ=00 channel, with the left hand cut ($l.h.c.$) obtained from
Eq.~(\ref{pade}), gives instead $M_\sigma=519$MeV and
$\Gamma_\sigma=579$MeV.} and the $\rho$ resonance in the IJ=11
channel: $M_\rho=708$MeV and $\Gamma_\rho=119$MeV which should be
compared with the experimental value $M_\rho=769$MeV and
$\Gamma_\rho=150$MeV. Actually, the quality of the Pad\'e
approximants can be much improved by tuning the $\bar l_i$
parameters within reasonable ranges, that is to fit the global
phase shifts  while the $\bar l_i$ parameters are still consistent
with the parameters constrained by low energy experiments. This
remarkable improvement is possible, as pointed out in
Refs.~\cite{GO99,OOR00}. Inspired by this approach (often called
unitarized chiral approach in the literature), we also made a
global fit to the experimental phase shifts using the [1,1] Pad\'e
amplitudes in Eq.~(\ref{pade}). In our fit the experimental data
are taken from Refs.~\cite{rho} -- \cite{Ro77}, especially we also
include the newest E865 Collaboration data from $K_{e4}$
decays~\cite{E865}. We fit the IJ=11 and 20 data up to
$\sqrt{s}=1GeV$ and the IJ=00 data up to
$\sqrt{s}=700$MeV.~\footnote{The data truncation in the IJ=00
channel is to reduce the coupled channel effects and the pollution
from $f_0(980)$.} The fit quality is impressive, as can be seen in
Fig~\ref{phase}.
%%%%%%%%%%%%%%%%%%%%%%%%%%%%%%%%%%%%%%%%%%%%%%%%%%%%%%%%%%%
%%%%%%%%%%%%%%%%%%%%%%%%%%%%%%%%%%%%%%%%%%%%%%%%%%%%%%%%%
The resulting  values of the $\bar l_i$ parameters corresponding
to the minimized $\chi^2$ are as follows, \be\label{our} \bar
l_1=-0.485\ , \,\,\, \bar l_2=5.696\ ,\,\,\, \bar l_3=16.136\
,\,\,\, \bar l_4=3.650\ . \ee Notice that in the above fit the
$\bar l_3$ parameter is much larger than that in Eq.~(\ref{sd})
and the value in Ref.~\cite{GO99}. This is not a serious  problem
as we found in the fit that though the minimization of the
$\chi^2$ prefers a large value of $\bar l_3$ but it is not
sensitive to $\bar l_3$. Another fact which is worth pointing out
is that the chiral estimate on $\bar l_3$ is crude and is subject
to large uncertainties~\cite{CGO01}. The physical resonances
obtained from our global fit are, $M_\rho=751$MeV,
$\Gamma_\rho=144$MeV and $M_\sigma=449$MeV,
$\Gamma_\sigma=482$MeV. The scattering length parameters in the
IJ=00 and 20 channels are found to be $a_0^0=0.194$ and
$a^2_0=-0.044$ which are reasonable as comparing with the
experimental results.

Though successfully predicting the existence of the $\sigma$ and
the $\rho$ resonances  qualitatively or even  quantitatively for
the latter, it is well known that the Pad\'e approximation
encounters severe problems in theory aspects. One of which is the
generation of dubious poles on the complex $s$ plane, as listed in
tables~1 and 2,~\footnote{The poles listed in tables~1 and 2 may
still be incomplete.} except for those physically accepted.
%%%%%%%%%%%%%%%%% TABLES %%%%%%%%%%%%%%%%%%%%%%%%%
\begin{table}[bt]
\label{tab1}
\centering\vspace{-0.cm}
\begin{tabular}{|c|c|c|c|c|c|}
\hline
&Pole status&Re[$s_{pole}$] (MeV) &Im[$s_{pole}$] (MeV) &residue ($GeV^2$)\\
\hline
IJ=11& $\rho$ &708 (M)&119 ($\Gamma$)&-0.07+0.15i
\\ \hline
      & $\sigma$ &430 (M) & 456 ($\Gamma$)& -0.19-0.23i
\\ \cline{2-5}
IJ=00 & BS$_1$ & $0.618m_\pi^2$ && $-5\times10^{-5}$
\\ \cline{2-5}
      & VS$_1$ & $0.621m_\pi^2$& &$6.2\times10^{-5}$
\\ \cline{2-5}
      & PSR & $-81.1m_\pi^2$& $60.6m_\pi^2$ &-4.76-4.50i
\\ \hline
      &BS$_1$  & $1.97528m_\pi^2$ & & $6\times 10^{-7}$
\\ \cline{2-5}
IJ=20  & VS$_1$&  $1.97525m_\pi^2$ & & $-6\times 10^{-7}$
\\ \cline{2-5}
     & VS$_2$ & $0.0495m_\pi^2$ & &$1.8\times10^{-3}$
\\ \cline{2-5}
      & PSR & $133.8m_\pi^2$& $483.4m_\pi^2$ & -26.27+4.09i
\\ \hline
\end{tabular}
\caption{ Resonances, physical sheet resonances (PSR), bound
states (BS) and virtual states (VS) as predicted by Pad\'e
approximation on the complex $s$ plane using the $\bar l_i$
parameters from Ref.~\protect\cite{GM}. The pole position
$s_{pole}=(M+i\Gamma/2)^2$. Here the residue of a second sheet
pole means the residue of the corresponding pole in $1/S$.}
\end{table}
\begin{table}[bt]
\label{tab3}
\centering\vspace{0.1cm}
\begin{tabular}{|c|c|c|c|c|c|}
\hline
&Pole status&Re[$s_{pole}$] (MeV) &Im[$s_{pole}$] (MeV) &residue ($GeV^2$)\\
\hline
IJ=11& $\rho$ &751 ($M$)&144 ($\Gamma$)&-0.10+0.19i
\\ \hline
      & $\sigma$ &449 ($M$) & 482 ($\Gamma$)& -0.16-0.27i
\\ \cline{2-5}
IJ=00 & BS$_1$ & $0.372m_\pi^2$ && $-7\times10^{-5}$
\\ \cline{2-5}
      & VS$_1$ & $0.376m_\pi^2$& &$8\times10^{-5}$
\\ \cline{2-5}
      & PSR & $-68.5m_\pi^2$& $47.1m_\pi^2$ & -3.56-3.55i
\\ \hline
      & BS$_1$  & $2.17832m_\pi^2$ & & $3\times 10^{-5}$
\\ \cline{2-5}
IJ=20  & VS$_1$ &  $2.17701m_\pi^2$ & & $-3\times 10^{-5}$
\\ \cline{2-5}
     & VS$_2$ & $0.0372m_\pi^2$ & &$1.4\times10^{-3}$
\\ \cline{2-5}
      & PSR & $106.0m_\pi^2$& $306.8m_\pi^2$ & -16.45+2.62i
\\ \hline
\end{tabular}
\caption{Resonances, physical sheet resonances (PSR), bound states
(BS) and virtual states (VS) as predicted by Pad\'e approximation
on the complex $s$ plane, using the $\bar l_i$ parameters obtained
from the global fit to the phases in the IJ=00,20,11 channels as
described in the text. The pole position
$s_{pole}=(M+i\Gamma/2)^2$. Here the residue of a second sheet
pole means the residue of the corresponding pole in $1/S$.}
\end{table}
%%%%%%%%%%%%%%%%%%%%%%%%%%%%%%%%%%%%%%%%%%%%%%%%%%%%
We categorize
these dubious poles into 3 classes: the first contains
the bound state pole
and the accompanying virtual state pole denoted as BS$_1$ and
VS$_1$ respectively (the latter corresponds to the zero in the
physical sheet below threshold) in the IJ=00,20 channels.
The positions of the pole and the accompanying
zero are very close to each other.
The second includes the resonance poles found in the physical sheet
(denoted as PSR), and the third class is for the virtual state (denoted as VS$_2$)
 near $s=0$
found in the IJ=20 channel.

There are two reasons to ignore the first class poles. The first
reason is that they have very tiny residues and can be safely
ignored numerically. The second reason is that they can be tuned
to disappear totally by varying  the $\bar l_i$ parameters in the
reasonable range. To see this more clearly we recall that the $S$
matrix of the [1,1] Pad\'e approximant takes the following
form,\footnote{In the following all terms should be understood as
their proper analytic continuation on the complex $s$ plane. For
example, ${\rm Im}T_4=\rho (T_2)^2$ and the $r.h.s.$ is well
defined on the complex $s$ plane.} \be
S^{[1,1]}=\frac{T_2-T_4^*}{T_2-T_4}=\frac{T_2-{\rm Re}T_4+i\rho
(T_2)^2} {T_2-{\rm Re}T_4-i\rho (T_2)^2}\ , \ee where the spin and
isospin indices are dropped for simplicity. A virtual state is an
$S$ matrix zero located on the real axis below the threshold which
requires $T_2=T_4^*$ whereas the bound state pole requires
$T_2=T_4$. Therefore the bound state pole and the virtual state
disappear simultaneously if the following requirements are met,
\be\label{cancel} {\rm Im}T_4=0\ ,\,\,\,\, {\rm Re}T_4=T_2\ . \ee
Since ${\rm Im}T_4=\rho (T_2)^2$ from perturbative unitarity the
first condition in Eq.~(\ref{cancel}) implies that the bound state
pole and the virtual state pole cancel each other when they move
towards the Adler zero position for the lowest order partial wave
amplitude. As a consequence, the Adler zero of the Pad\'e
approximant is still there which coincides with the one of the
lowest order partial wave amplitude, but the order of the zero
decreases from two  to one (the second order Adler zero is
unnecessary from theoretical point of view). The second condition
of Eq.~(\ref{cancel}), which now reads ${\rm Re}T_4=0$, affords a
constraint that those $\bar l_i$ parameters have to
obey.~\footnote{Explicitly it is \be\label{20} 2(40+32 \bar l_1+32
\bar l_2-36 \bar l_3) +28 \pi-21 {\pi}^2=0 \ee for IJ=20, and
\be\label{00} 8141+1498 \bar l_1+2212 \bar l_2-1260 \bar
l_3+454i\sqrt{7}\ln({{7i+\sqrt{7}} \over {-7i+\sqrt{7}}})
+1056{\ln^2({{7i+\sqrt{7}} \over {-7i+\sqrt{7}}})}=0 \ee for
IJ=00.} To cancel the bound state in the IJ=20 channel, we can for
example choose the following set of $\bar l_i$: $\bar
l_1=-0.5,\bar l_2=5.3,\bar l_3=3.7,\bar l_4=4.0$, satisfying
Eq.~(\ref{20}) numerically which fit the experimental values of
$m_\rho$ and $\Gamma_\rho$ well. For these $\bar l_i$ parameters
the experimental phase shifts of three channels IJ=00,20 and 11
are also fitted well. Similar situation happens in the IJ=00
channel. But when we try to cancel the bound states both in IJ=00
and IJ=20 channels simultaneously, it is difficult to find a  set
of $\bar l_i$ parameters within reasonable ranges to agree well
with the experimental phase shifts simultaneously in three
channels. We ascribe this difficulty to the non-perfectness of the
Pad\'e approximation itself.

\begin{figure}[htb]
\begin{center}
\mbox{\epsfxsize=65mm \epsffile{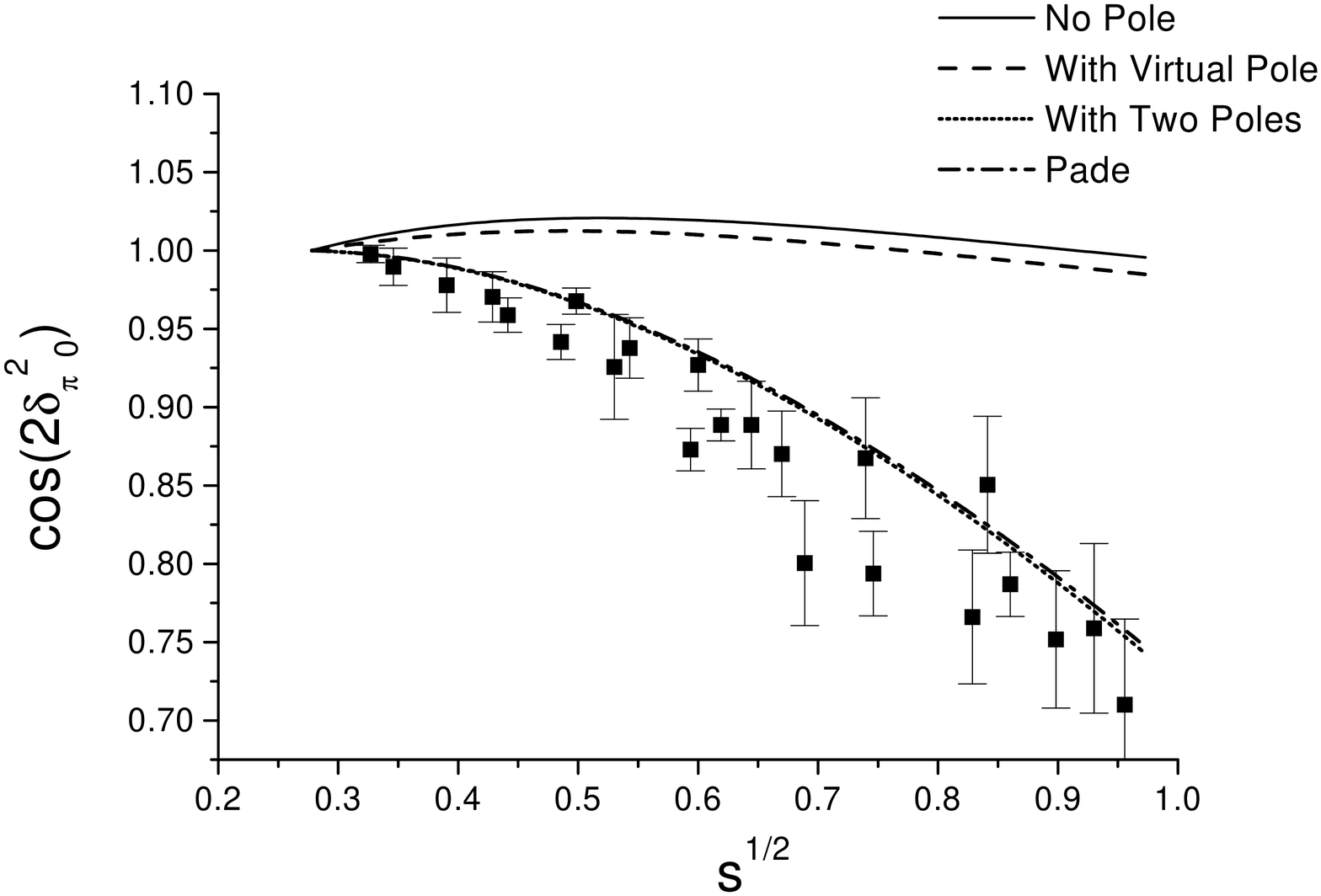}} \hspace*{10mm}
\mbox{\epsfxsize=65mm \epsffile{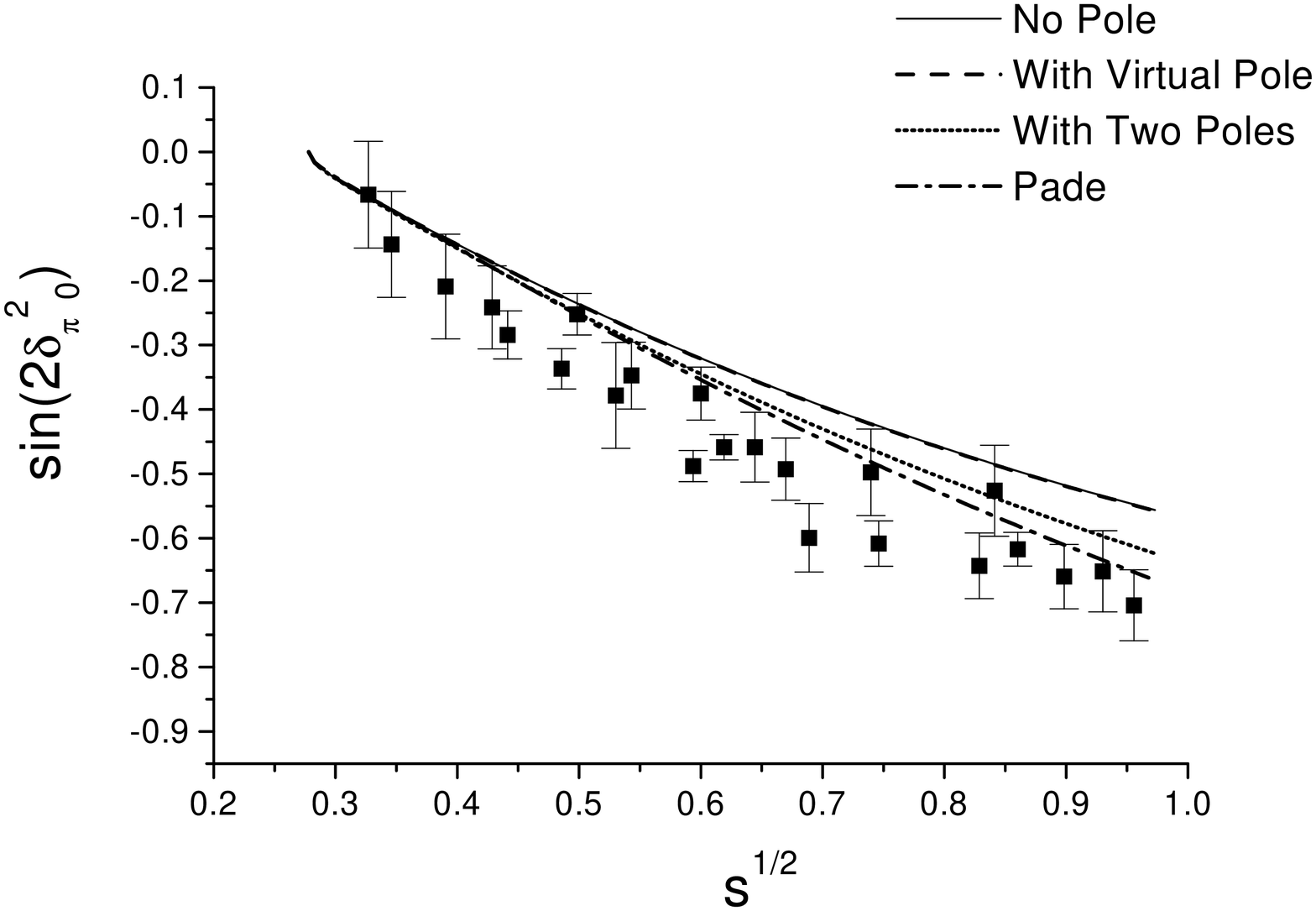}} \vspace*{0mm}
\caption{ \label{cos2d20f}Various contributions to
$\cos(2\delta^2_{\pi 0})$ and $\sin(2\delta^2_{\pi 0})$ in the
IJ=20 channel. Solid lines represent only the left hand integral
contributions in Eq.~(\ref{cos2d20}) and Eq.~(\ref{sin2d20}). In
the dashed line the virtual state (VS$_2$) contribution is added;
in the dotted line the contributions from both the virtual state
and the physical sheet resonance are added. The dotted line
numerically coincides with the dot-dashed line as it should be,
the latter is obtained by evaluating the Pad\'e amplitude in the
physical region, using the $\bar l_i$ parameters from
Ref.~\protect\cite{GM}.}
\end{center}
\end{figure}

The existence of the physical sheet resonances violate causality
and must be a false prediction of Pad\'e approximation. One may
argue that such kind of false poles locate very far from the
physical region and the region where ChPT is valid (inside the
circle $|s|<<1$GeV$^2$), and therefore the Pad\'e approximation is
still  acceptable numerically in phenomenology. However a careful
analysis just reveals the opposite. As can be seen from tables~1
and 2 the residues of the PSRs are usually very large that in some
cases it strongly plagues the prediction of the Pad\'e
approximation.  To see this more clearly, taking the IJ=20 channel
for example, we recall the following dispersion
relations~\cite{XZ00,XZ01}:
 \bqa \cos(2\delta_{\pi 0}^2)
&=&1+{(s-4m_\pi^2)\over2\pi}\int_{-\infty}^0 {{\rm Im}
(S(s')+{1\over S(s')})\over (s'-s)(s'-4m_\pi^2)} ds' ,
\label{cos2d20}
\\ \sin(2\delta_{\pi 0}^2)&=&
 \rho(s)\left[2a^2_0+{(s-4m_\pi^2)\over 2\pi}\int_{-\infty}^0{{\rm Im}
 ({1\over i\rho} (S(s')-{1\over
 S(s')}))\over (s'-s)(s'-4m^2_\pi)} ds' \right]. \label{sin2d20}
\eqa
 These two equations  are obtained by assuming no pole exists
on both sheets. But pole contributions from different sheets can
be added to the above equations easily. The left hand integrals
can be evaluated by using the expressions obtained from the Pad\'e
approximant $T^2_0$. The left hand side of  Eqs.~(\ref{cos2d20})
and (\ref{sin2d20}) can be directly evaluated from the Pad\'e
amplitudes in the physical region. The right hand side of
Eqs.~(\ref{cos2d20}) and (\ref{sin2d20}) separate different type
of contributions, i.e., from cuts, resonances, bound states,
virtual states or physical sheet resonances. It is clearly seen in
Fig.~\ref{cos2d20f}, that the contribution from the left hand
integral in Eq.~(\ref{cos2d20}) deviates  far from the
experimental value. Only after adding the physical sheet resonance
in Eq.~(\ref{cos2d20}) may we reproduce the phase in the physical
region.  Though in Eq.~(\ref{sin2d20}) the left hand integral
works rather well numerically, the predictions of Pad\'e
approximants on poles and cuts become no longer trustworthy
quantitatively, in the situation when there exist physical sheet
resonances with large residues.
%%%%%%%%%%%%%%%%%%%%%%%%%%%%%%%%%%%%%%%%%%%%%%%%%%%
%%%%%%%%%%%%%%%%%%%%%%%%%%%%%%%%%%%%%%%%%%%%%%%%%%%

\begin{figure}[hbtp]
\begin{center}
\vspace*{0mm}
\epsfysize=70mm
\epsfbox{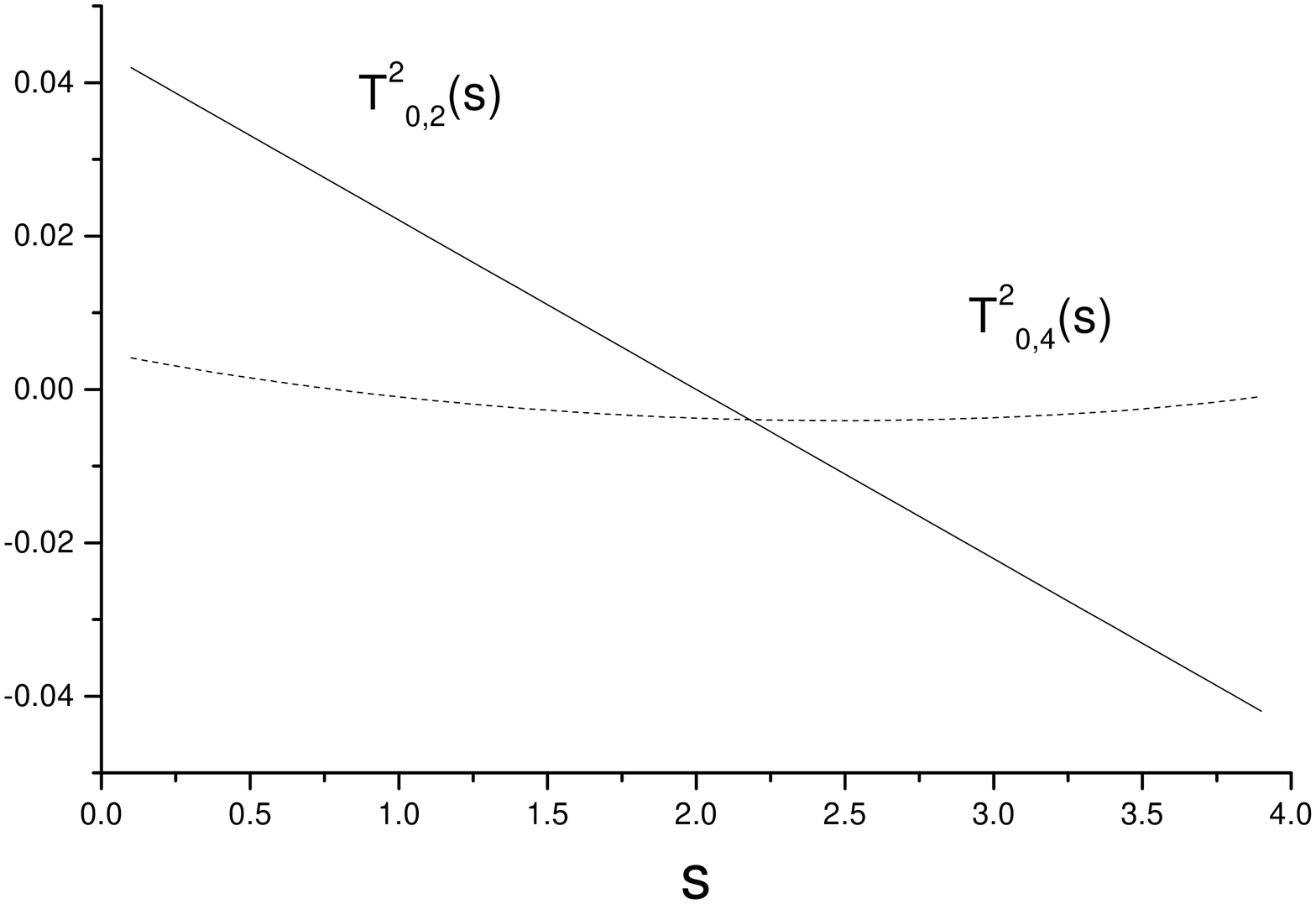}
\vspace*{5mm} \caption{ \label{t24}$T^2_{0,2}$ and $T^2_{0,4}$ amplitudes
using $l_i$ parameters obtained from
our global fit to the Pad\'e amplitudes in IJ=00, 20 and 11 channels. The  bound
state pole locates at the point where the two amplitudes meet.
}
\end{center}
\end{figure}

From the above discussions we realize that the existence of the
physical sheet resonance plagues the predictive power of the
Pad\'e approximant $T^2_0$, at quantitative level. Similar things
happen also in the IJ=00 channel.~\footnote{In the IJ=00 channel
the PSR contribution to $\sin(2\delta_\pi)$ becomes also sizable.}
Fortunately Pad\'e approximants still work qualitatively in the
sense that it correctly predicts the existence of the physical
resonances, like $\sigma$ and $\rho$. In here it is worth
emphasizing that the virtual state VS$_2$ in the IJ=20 channel may
be physical. The residue of such pole, though very small, is much
larger than the residues of BS$_1$ and VS$_1$, and it improves the
$r.h.s.$ of Eq.~(\ref{cos2d20}). There is a simple reason for the
existence of such a pole which does not rely on the detailed form
of the Pad\'e approximant $T^2_0$. To understand this, we recall
that  around the bound state pole at $s=s_B$ the $S$ matrix can be
expanded as $r/(s-s_B) +C + O((s-s_B))$ where $C$ is a constant.
The accompanying  zero will occur at $s_0\simeq s_B-r/C$ for any
non zero value of $C$. This simple mechanism explains the pair
production of BS$_1$ and VS$_1$. The reason for the generation of
VS$_2$ is similar. Remember that $S=1+2i\rho T$ and $i\rho$ is
purely real and negative when $0<s<4m_\pi^2$. When $s\rightarrow
0$ from positive side $i\rho\rightarrow -\infty$ therefore if
$T(0_+)$ is positive then $S(0_+)\rightarrow -\infty$. But since
$S(4m_\pi^2)=1$ therefore the $S$ matrix must go through a zero
(at least once) located somewhere between $s=0+$ and $s=4m_\pi^2$.
It happens that the [1,1] Pad\'e approximation gives a negative
value to $T_0^{0[1,1]}(0_+)$ and $T_1^{1[1,1]}(0_+)$, but a
positive value to $T_0^{2[1,1]}(0_+)$. From this analysis we
realize that the existence of the virtual state VS$_2$ is a
consequence of more general situation ($T(0_+)> 0$) than that of
the concrete form of the Pad\'e approximant $T^{2[1,1]}_0$. In
here the signs of the Pad\'e amplitudes $T^{0[1,1]}_0(0_+)$,
$T^{1[1,1]}_1(0_+)$ and $T^{2[1,1]}_0(0_+)$ coincide with the
signs of the lowest order amplitudes $T^0_{0,2}(0_+)$,
$T^1_{1,2}(0_+)$ and $T^2_{0,2}(0_+)$, respectively, and the
lowest order amplitudes are unambiguous predictions from current
algebra. The coincidence of the sign simply reflects the fact that
the lowest order amplitudes dominate at $s\sim 0$, as can be
clearly seen in Fig.~\ref{t24}. Though the virtual state may
really exist, its effect may be very small due to its small
coupling, as indicated by tables~1 and 2.

To conclude, the dispersion relations set up in Refs.~\cite{XZ00}
and \cite{XZ01} enable us to examine explicitly contributions from
different types of dynamical singularities -- the resonances, the
left--hand cuts, and the bound states or virtual states -- to the
phase shifts. Hence a critical examination on  the Pad\'e
approximation becomes possible. We find that even though the
Pad\'e approximation can give a reasonable global fit to the
$\pi\pi$ scattering phase shifts, the contribution to the phase
shifts can be largely from disastrous physical sheet resonances in
some situations. In such cases the other predictions of the Pad\'e
amplitudes become unreliable either, at least at quantitative
level. However we argue that in the IJ=20 channel the virtual
state close to $s=0$ as predicted by the Pad\'e approximation is a
consequence of more general conditions and seems to exist
physically.
%%%%%%%%%%%%%%%%%%%%%%%%%%%%%%%%%%%%%%%%%

%%%%%%%%%%%%%%%%%%%%%%%%%%%%%%%%%%%%%%%%%%%%%%%%%%%%%%%%%%%%%%%%%
%%%%%%%%%%%%%%%%%%%%%%%%%%%%%%%%%%%%%%%%%%%%%%%%%%%%%%%%%%%
\end{document}